\documentclass[12pt,preprint]{aastex}
\usepackage{rotating}
\def\msun{\rm M_{\sun}}

\def\av{${\rm A_V}$}

\begin{document}

\shortauthors{Hernandez et al.}
\shorttitle{Debris Disk in Orion OB1}

\title{Spitzer observations of the Orion OB1 association: second generation dust
disks at 5-10 Myr}

\author{Jes\'us Hern\'andez\altaffilmark{1,2},  
C\'esar Brice\~no\altaffilmark{2}, Nuria Calvet\altaffilmark{1}, Lee Hartmann\altaffilmark{1}, 
James Muzerolle\altaffilmark{3} and Amilkar Quintero\altaffilmark{4,5}}

\altaffiltext{1}{Department of Astronomy, University of Michigan, 830 Dennison Building, 500 Church Street, Ann Arbor, MI 48109, US}

\altaffiltext{2}{Centro de Investigaciones de Astronom\'{\i}a, Apdo. Postal 264, M\'{e}rida 5101-A, Venezuela.}

\altaffiltext{3}{Steward Observatory, University of Arizona, 933 North Cherry Avenue, Tucson, AZ 85721, US}

\altaffiltext{4}{Universidad de Carabobo, FACYT, Dept. de F\'{\i}sica, Venezuela}

\altaffiltext{5}{Visiting student at CIDA}

\email{jesush@cida.ve}

\begin{abstract}
 We report new Spitzer observations of intermediate mass stars in two regions of the 
Orion OB1 association located in the subassociations OB1a ($\sim$10 Myr) 
and OB1b ($\sim$5 Myr). In a representative sample of stars earlier than F5  
of both stellar groups, we find a population of stars surrounded 
of debris disks, without excess in the IRAC bands and without emission 
lines in their optical spectra, but with a varying degree of 24{\micron} excess. 
Comparing our samples with 24{\micron} observations of intermediate mass stars 
in other stellar groups, spanning a range of ages from 2.5 Myr to 150 Myr, 
we find that debris disks are more frequent and have 
larger 24{\micron} excess at 10 Myr (OB1a). 
This trend agrees with predictions of models of evolution of solids in the 
outer regions of disks ($>$30 AU), where large icy objects 
($\sim$1000 Km) begin to form at $\sim$10 Myr; the presence of these objects in the disk
initiates a collisional cascade, producing enough dust particles to explain the relatively 
large 24 {\micron} excess observed in OB1a.  
The dust luminosity observed in the stellar groups older than 10 Myr declines 
roughly as predicted by collisional cascade models.  
Combining Spitzer observations, optical spectra and 2MASS data, 
we found a new Herbig Ae/Be star (HD290543) and a star (HD36444) with a large
24 {\micron} excess, both in OB1b. This last object could be explained as a
intermediate stage between HAeBe and true debris systems or as a massive debris 
disk produced by a collision between two large objects ($>$1000 Km).
\end{abstract}


\section{Introduction}
\label{sec:int}
Theories of star formation indicate that, in general,
stars are born surrounded by disks due to angular
momentum conservation \citep{hartmann05}. These optically 
thick primordial disks, which contain gas and dust, are 
expected to evolve by accreting gas into the star and
planets, while dust grains grow and settle towards the
mid-plane of the disk. 
As the primordial disk evolves from optically thick to optically thin, 
its excess at near and mid infrared wavelength diminishes drastically. 
The time scale for this evolution is strongly dependent on 
the stellar mass \citep{lada95,muzerolle03,calvet04,aurora05}; in particular,
for low mass stars (K5 or later), $\sim$90\% of the stars have lost
their primordial disk at about 5-7 Myr \citep{haisch01, hillenbrand05}, 
while for objects in the mass range of the Herbig Ae/Be (HAeBe) stars
($\sim 2 - 8 \msun$), the time scale for primordial disk dissipation 
is less than $\sim$3 Myr \citep{hernandez05}. 
Theories of dust evolution in the solar nebula indicate that the timescale 
for disk evolution is also proportional to the orbital period 
\citep{weidenschilling97, dullemond04}; so, grain growth and settling to 
the mid-plane occur fastest in the inner disk resulting in a faster decline 
of disk emission at shorter wavelength \citep{hartmann05,aurora06, lada06}. 
In this ''clearing phase'', 
stars could evolve to ''transition disk objects'' in which we find an inner 
optically thin region with an outer optically thick primordial disk 
\citep{calvet02, paola05, calvet05}. Since a small fraction of
transition objects are observed in several star formation regions 
\citep[e.g., ][]{aurora06,muzerolle06}, 
this phase has to be very  brief \citep{kb04a}. 
After this brief phase, second generation
dusty disks (debris disks) are frequently observed at mid and far IR 
wavelength in stars with age from few Myr to several Gyr 
\citep[e.g., ][]{rieke04, decin03, bryden06, chen05a, chen05b}.
Since radiation pressure, Poynting-Robertson drag, ice sublimation and other processes
remove the dust on short timescales compared with the age of the system, 
the observed dust must have been replenished from a reservoir, such as sublimation of
comets or collisions between parent bodies \citep[e.g., ][]{chen06,  kb04b, dominik03}.
In the collisions scenario, models
of debris disk evolution suggest that the formation of  disks coincides
with the formation of large icy objects (1000 Km) at 10-30 Myr, 
which stir up the leftover objects in the disks
originating a collisional cascade that produces a copious amount of dust
observed in young debris disk stars \citep{kb05, kb04a, kb04b,  decin03}. 
For A-type stars,  \citet{kb02,kb04b} and \citet{dominik03} 
show that, as the debris reservoir diminishes and the dust is removed,
the luminosities of the debris disks decay in several hundred millions years, 
following a simple exponential law, age$^{-1}$. 
Observations confirm  this long period trend \citep{decin03, rieke05}. 
HAeBe stars are the  precursors of A-type stars
with debris disks, like Vega or $\beta$ Pic. However, details of the
earlier processes  by which the primordial disk evolves to transition
disk and to debris disk are not well understood. 

In this contribution, we address the questions of
disk dissipation and debris disk formation in the
mass range of HAeBe. We present Spitzer space telescope
data for intermediate mass stars located in two regions with different
evolutionary stages in the Orion OB1 association,
one of the largest and nearest regions with active star
formation. The selection of the samples and observations
are described in \S2. We analyze the observations and describe the
results on \S3 and give our conclusions in \S4.

\section{Observations and sample selection}
\label{sec:obs}

\begin{figure}
\includegraphics[angle=270,scale=0.60]{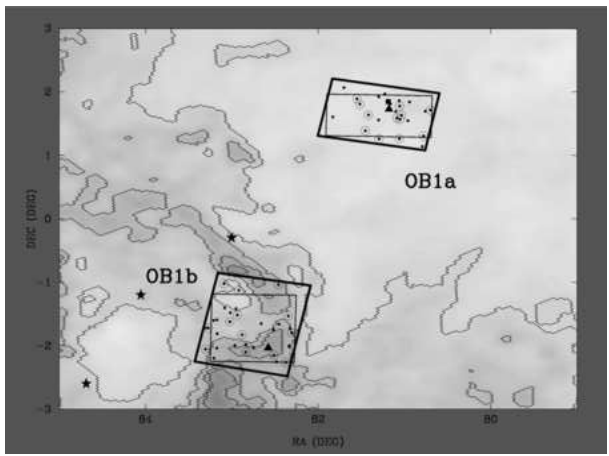}
\caption{Observations in Orion OB1 association. The isocontours of the map of dust
infrared emission \citep{schlegel98} are indicators of galactic extinction
(\av=0.2, 0.4, 0.6 \& 0.8 mag). 
The Ori OB1b sub-association corresponds mostly to the belt region,
and can be traced by the ringlike structure centering at RA$\sim$84
and DEC$\sim$-2, whereas Ori OB1a spans the large low reddening area north and west
of Ori OB1b.
The thick boxes represent the MIPS fields and the thin boxes are the
IRAC fields. Dots are the selected early type stars
in each field. Open circles are objects with 24 {\micron} excess (see \S 3). These stars,
and stars indicated with
a square (25 Ori in OB1a), and triangles (V346 Ori in OB1a and HD290543 in OB1b)
have also been observed with the FAST spectrograph.}
\label{fig:field}
\end{figure}

We have obtained near-infrared (NIR) and mid-infrared photometry  of 
two regions in the Orion OB1 association using the 24 {\micron} band 
of the Multiband Imaging Spectrometer for Spitzer \citep[MIPS;][]{rieke04}
and the four channels (3.6, 4.5, 5.8 \& 8.0 {\micron}) of the InfraRed 
Array Camera \citep[IRAC, ][]{fazio04}, instruments on board the Spitzer 
Space Telescope. Figure \ref{fig:field} shows the positions of the MIPS (thick solid line)  
and IRAC\footnote{Figure \ref{fig:field} shows the IRAC bands 1 and 3, 
the bands 2 and 4 have a 8' displacement to the north} 
(thin solid line) fields in the Orion OB1 association. One of these fields is located 
on the Orion OB1a sub-association with an age of 10 Myr \citep{briceno06} 
and the other field is located on the Orion OB1b sub-association with an age of 5 Myr 
\citep{briceno05}. Using isocontours of the dust infrared emission 
map \citep{schlegel98}  we can estimate that at least 90\% of the area in OB1a has visual 
extinctions smaller than {\av}$\sim$0.12 and at least 90\% of the area in OB1b has 
 visual extinctions smaller than {\av}$\sim$0.6. Table 1 includes more information 
about the MIPS fields and the related stellar groups. 

MIPS observations were done using medium scan mode with full-array
cross-scan overlap, resulting in a total effective exposure time per pointing
of 40 seconds.  The images were processed using
the MIPS instrument team Data Analysis Tool (DAT), which calibrates
the data and applies a distortion correction to each individual exposure
before combining into a final mosaic \citep{gordon05}.  We also applied
an additional correction to remove a faint readout-dependent residual
pattern.
We obtained point source photometry at 24 {\micron} with IRAF/daophot 
point spread function fitting, using an aperture size of about 5.7" and
an aperture correction factor of 1.73 derived from the STinyTim PSF model.
The absolute flux calibration uncertainty is less than 5\%.
Our final flux measurements are complete down to about 1 mJy in both maps.

The IRAC observations were done using a standard raster map with 290" offsets,
to provide maximum areal coverage with just a slight overlap between frames,
to aid in mosaicking the data.  Each position is composed of 3 dithers.
The single-frame integration was 12 seconds.
The IRAC Basic Calibrated Data (BCD) were processed using the 
{\sl IRACproc} \citep{schuster06} package. IRACproc improves 
cosmic ray and other transient rejection by using spatial derivative images to 
map the locations and structure of astronomical sources. Because the native IRAC
images are under/critically sampled (1.22 arcsec/pixel), 
the PSF is subject to large variations
in shape between successive frames because of sub-pixel shifts caused by
dithering or telescope jitter. The software is designed to preserve the 
photometric integrity of the data, especially of bright sources, by applying 
a metric that accounts for these large variations in the PSF.
Although IRACproc was originally developed for the Nearby Stars Guaranteed Time
Observer (GTO) program, which was based on a 5-point small scale dither pattern,
our IRAC 3-point dither observations were sufficient to provide a reliable
cosmic ray/transient rejection.  The final mosaics created with IRACproc
have a scale of 0.86 arcsec/pixel. We extracted the photometry
using the {\sl apphot} package in IRAF, with an
aperture radius of 12 arcsec and a background annulus 
from 12 to 22.4 arcsec. Because IRAC standard stars were measured with the same
aperture and sky annulus, we did not apply an aperture correction.
We adopted zero-point values of 19.660, 18.944, 16.880, 17.394 in 
the [3.6], [4.5], [5.8] and [8] bands, respectively \citep{hartmann05}.
Since MIPS and IRAC fields do not cover the same area on the sky (see Figure \ref{fig:field}), 
some objects were not observed in all IRAC channels. However, 
the results shown in \S 3 are based mainly on the MIPS data.

\begin{figure}
\epsscale{0.7}
\plotone{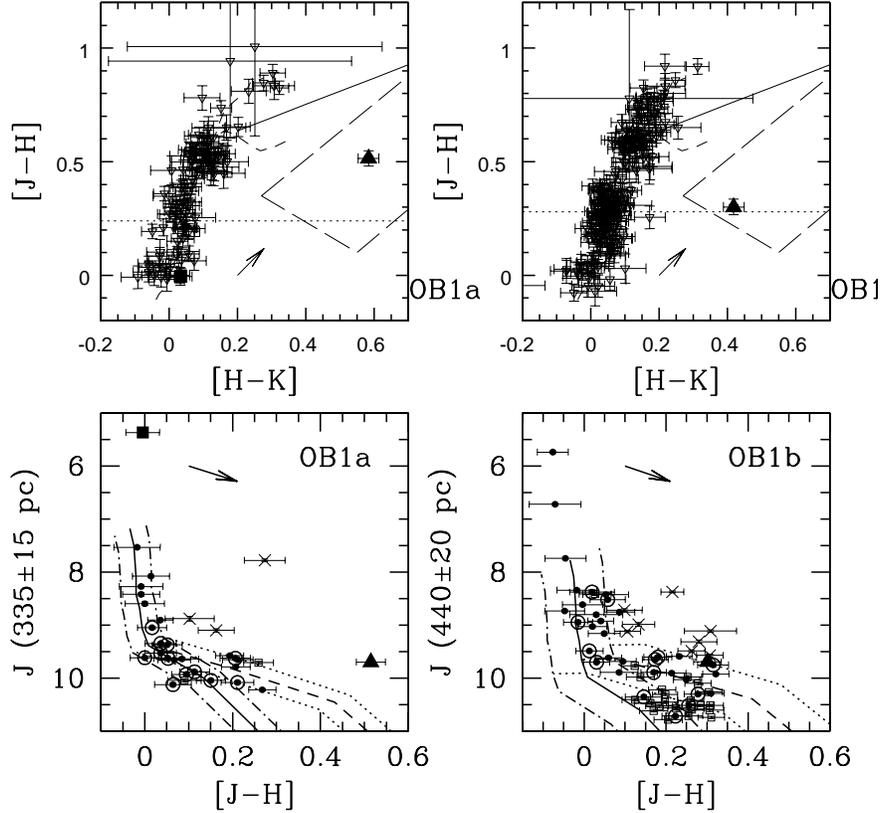}
\caption{NIR color-color and color-magnitude diagrams illustrating the procedure 
for sample selection. The upper panels show the NIR color-color 
diagrams for stars with J$<$10.3 for the OB1a field (left) and J$<$11.0 
for the OB1b field (right); the dotted lines represent the limits in 
[J-H] color, 0.24 for OB1a and 0.28 for OB1b (see text). 
Most of the stars (inverse triangles),  
including the star 25 Ori (filled square),  are located 
on the main sequence locus \citep[dashed line; ][]{bessell88}. Only 
two objects (V346 Ori in OB1a and HD290543 in OB1b; filled triangles) appear in the HAeBe 
stars loci \citep[long-dashed line; ][]{hernandez05}. By comparison the 
locus of low mass stars with primordial disks (CTTs) is indicated with a  
solid line.  These HAeBe stars and the early type objects without NIR excess 
(below the [J-H] limits) are plotted in the color-magnitude diagrams, J versus [J-H] (lower panels). 
The solid lines and the dashed lines represent the ZAMS and the isochrones 
for the distance and the age of each stellar group. The dot-dashed lines 
(for the ZAMS) and the dotted lines (for the isochrones) show the 
regions expected when the extinctions and the errors of the distances 
(see Table 1) are included in the evolutionary models. Symbol X represents 
foreground objects located above these regions. Open squares are 
objects without MIPS detections. Filled circles represent stars selected in each 
sub-association, and the stars with 24{\micron} excess are surrounded 
by open circles. Since 7 stars in OB1a and 14 stars in OB1b have large intrinsic 
proper motion, they are not include in the color-magnitude diagrams. 
Errors bars represent the uncertainties from 2MASS catalog. 
and the arrows are the reddening vectors for \av=1.  }
\label{fig:jhk}
\end{figure}

To find the early type stars (F5 or earlier) in both fields,
we selected from the 2MASS catalog \citep{cutri03} stars with J$<$10.3 for 
the OB1a field and J$<$11.0 for the OB1b field. These limits were calculated 
using the V magnitude \citep{allen2000} and the V-J color \citep{kh95} for a 
main sequence star with spectral type F5, and the  
distances and the visual extinctions in Table 1. Using the \citet{ccm89} extinction law with 
R${\rm _V}$=3.1,  we transformed visual extinctions to reddening in the NIR bands. 
The NIR color color diagrams, [J-H] versus [H-K] (upper panels of Figure \ref{fig:jhk}), 
for the stars selected in both fields reveal 
that most of the stars are located on the main sequence locus \citep{bessell88}. Therefore, 
these stars have little extinction and small or no NIR excesses. Only two objects, V346 Ori in OB1a 
and HD290543 in OB1b, appear in the HAeBe stars loci \citep{hernandez05}, 
where the NIR excess can be explained by  emission from an optically thick primordial disk 
with a sharp dust-gas transition at the dust destruction radius 
\citep[e.g., ][]{dullemond04, muzerolle04}. 
V346 Ori and HD290543 are included as members of OB1a and OB1b, respectively.
The dotted horizontal lines in the upper panels of Figure \ref{fig:jhk} are 
the [J-H] limits corresponding to a F5 star with \av=0.12 for OB1a 
and to a F5 star with \av=0.6 for OB1b. The lower panels of Figure \ref{fig:jhk}
show the color-magnitude diagrams, [J-H] versus J, for the stars below the [J-H] limits in the upper panels and 
the stars with thick primordial disk (solid triangles). The solid lines represent the Zero Age Main Sequence (ZAMS)
from \citet{palla93} at the distance of each stellar group \citep{hernandez05}; dashed lines represent 
the 10 Myr and 5 Myr isochrones \citep{palla93} for OB1a and OB1b, respectively. Dot-dashed lines (for the ZAMS) 
and dotted lines (for the isochrones) show the region expected when the extinctions and the errors of the distances 
are included in the evolutionary models. Stars above these regions are foreground objects (X's), 
so we rejected them as members of OB1a and OB1b. 
Since the motion of the Ori OB1 association is
mostly directed radially away from the Sun, the expected
intrinsic proper motions in right ascension
and declination are small.  
Using the proper motions in \citet{kharchenko01},
we rejected 7 stars in OB1a and 14 stars in OB1b, 
with a relative large intrinsic 
proper motions, applying  the criterion defined by \citet{brown99}.
For the leftover objects (likely members), 93\% of the stars in OB1a and 62\% of the stars in OB1b have 
MIPS detections; most stars without [24] measurements (open square) exhibit 2MASS photometry characteristic of stars 
later than F0. Comparing 2MASS sources with and without MIPS detections in each stellar group,
we find that the samples are complete for stars with J$\leqq$10. Using the distances, 
ages and reddening in Table 1 we estimate that this completeness limit corresponds roughly to a F0 stars 
in OB1a and to a A8 star in OB1b. However, since we detect stars with and without 24 {\micron}
excess in our faintest limits (F5), the selected stars are representative samples 
of intermediate mass stars in both fields.  
Our final samples of early type stars include 26 objects in OB1a field 
and 34 objects in Orion OB1b field.

We find that 28 of the stars in our sample show excesses at 24{\micron} 
above the photospheric colors (\S 3); optical spectra were obtained for 
these stars using the 1.5 m telescope of 
Whipple Observatory with the FAST Spectrograph \citep{fabricant98},
equipped with the Loral 512 x 2688 CCD. The spectrograph was set up in the FAST standard
configuration (3400{\AA} of spectral coverage centered at 5500\AA, with a resolution of $\sim$6\AA).
These data set were used for spectral classification following the methods described in \citet{hernandez04} and 
for discriminating stars with and without emission lines. In Figure \ref{fig:spec} we show
FAST spectra for the three emission line stars in our sample: the classical Be (CBe) star 25 Ori 
\citep{yudin01,hernandez05,briceno06}, 
and the two HAeBe stars (V346 Ori and HD290543). We also show the spectrum for the star HD36444 
which does not exhibit emission lines and has a large 24{\micron} excess (see \S \ref{sec:res}).

Table 2 shows spectroscopic \& photometric data for the intermediate and high mass members found in OB1a and OB1b.
Columns 1 and 2 show the 2MASS denomination and other names of the stars. Columns 3, 4, 
5 and 6 show the IRAC magnitudes for the sample. The typical errors
in these values are dominated by the uncertainties in the zero points
magnitudes, 0.02-0.05 mag \citep{hartmann05}. Columns 7 and 8 show [24]
MIPS data in magnitudes and fluxes, respectively. The typical errors 
in MIPS photometry are 0.03-0.06 mag. Column 9 shows the excess ratio of the measured 
flux at 24 {\micron} to that expected from the stellar photosphere (see next section). 
Spectral types and their references are shown in columns 10 and 11. The last column shows 
the disk type around each star (see next section).

\section{Results}
\label{sec:res}
\subsection{Observational analysis}
\begin{figure}
\epsscale{0.9}
\plotone{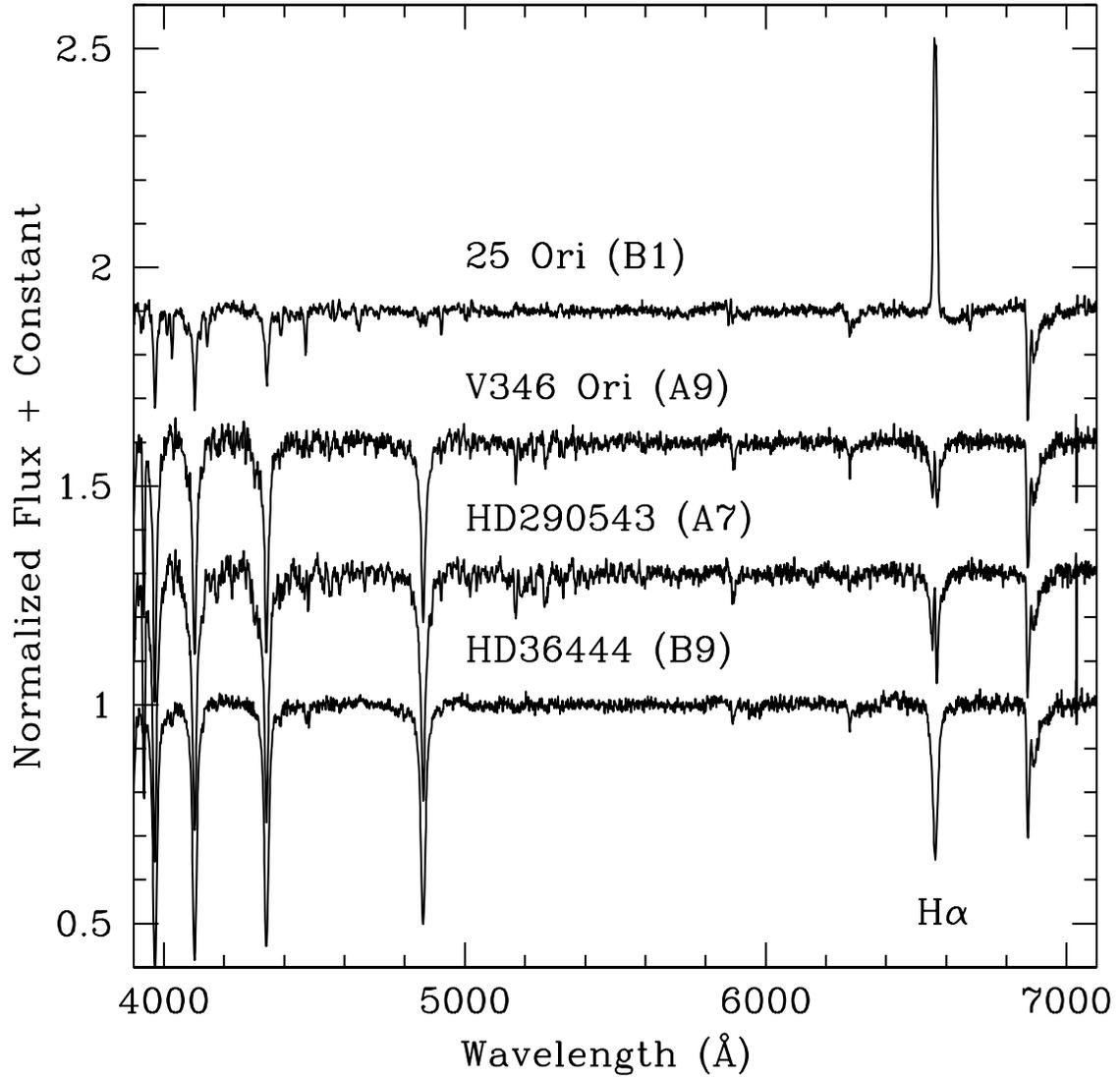}
\caption{FAST spectra of 25 Ori, V346 Ori, HD290543, and HD36444. The three first 
spectra are the emission line stars ( one CBe and two HAeBe) that we found in our data set. 
The last one is a star with large 24{\micron} excess, which does not exhibit emission features 
in its spectrum.}
\label{fig:spec}
\end{figure}

\begin{figure}
\epsscale{0.9}
\plotone{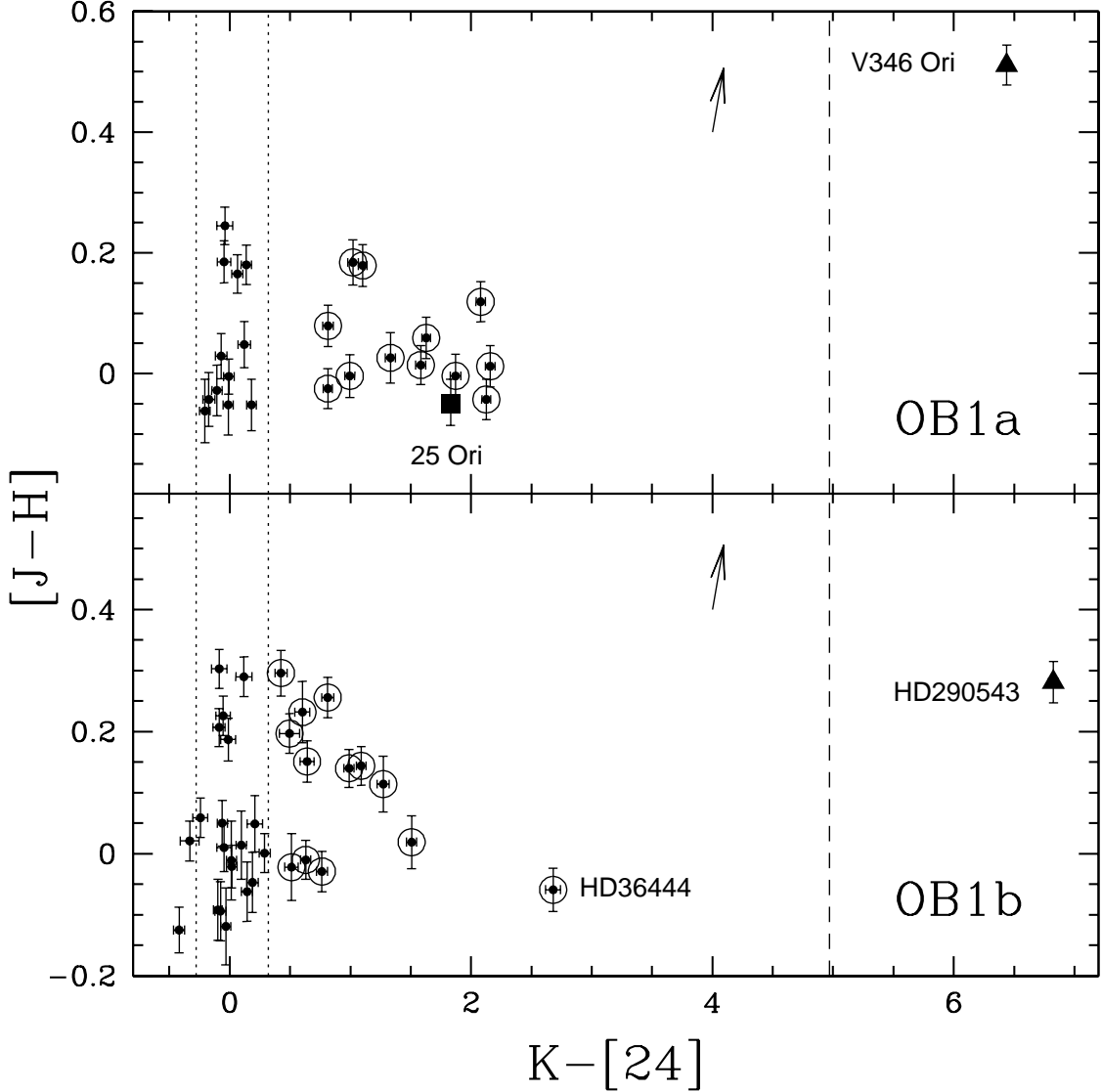}
\caption{
Location of the early type stars in the
[J-H] versus K-[24] color-color diagrams. 
The upper panel corresponds to the OB1a field and 
the lower panel to the OB1b field. Symbols are 
the same as Figure \ref{fig:field}. Arrows are the reddening vectors for 
\av=1.
The dotted lines represent the main sequence limits defined 
using the 3 $\sigma$ boundaries ($\sigma$=0.10, mean=0.02) 
on the K-[24] distribution of stellar sources compiled 
by \citet{kharchenko01} for the low reddening OB1a field. 
The dashed line at K-[24]$\sim$5 separates the debris disk/CBe region
from the HAeBe region and corresponds to the color of the star HR4796A.
}
\label{fig:mips}
\end{figure}

Figure \ref{fig:mips} shows the color-color diagram, [J-H] versus K-[24], for the early type stars 
selected as described in \S 2. Arrows represent the reddening vector for \av=1. 
There are clearly three regions in the diagrams. 
The first one is the main sequence region, 
which is defined using the K-[24] distribution of the 
stellar sources  compiled by \citet{kharchenko01} on the 
low reddening region OB1a. This distribution describes 
a Gaussian centered at K-[24]=0.02 with 
$\sigma$=0.10 mag,  and represents the 3 $\sigma$ 
boundaries on non-excess stars at 24 {\micron} (dotted lines). 
The second one is the region 
located between the main sequence region and the color of the
star HR4796A (K-[24]$\sim$5.0; dashed lines), 
which is the star with the largest 24 {\micron} excess in the data set 
of debris disk objects compiled by \citet{rieke05};
stars with optically thin disks like Vega type stars are located in this region 
\citep[][; hereafter debris disk region]{gorlova04,young04}. 
The CBe star 25 Ori (filled square) in OB1a, with small NIR excess
(upper-left panel in Figure \ref{fig:jhk}) and with H$\alpha$  in emission (Figure \ref{fig:spec}), 
is also located in this region. In this case the 24 {\micron} excess originates 
in a gaseous envelope, in which the emission lines and the NIR continuum excess 
are produced by free-free emission process.
The third region is where stars with strong infrared excesses due to 
optically thick disks, that is the HAeBe stars, are located.  
The well known HAeBe star V346 Ori \citep[Figure \ref{fig:spec}; ][]{hernandez05, vieira03, the94}
appears in this last region (filled triangle).

\begin{figure}
\epsscale{0.9}
\plotone{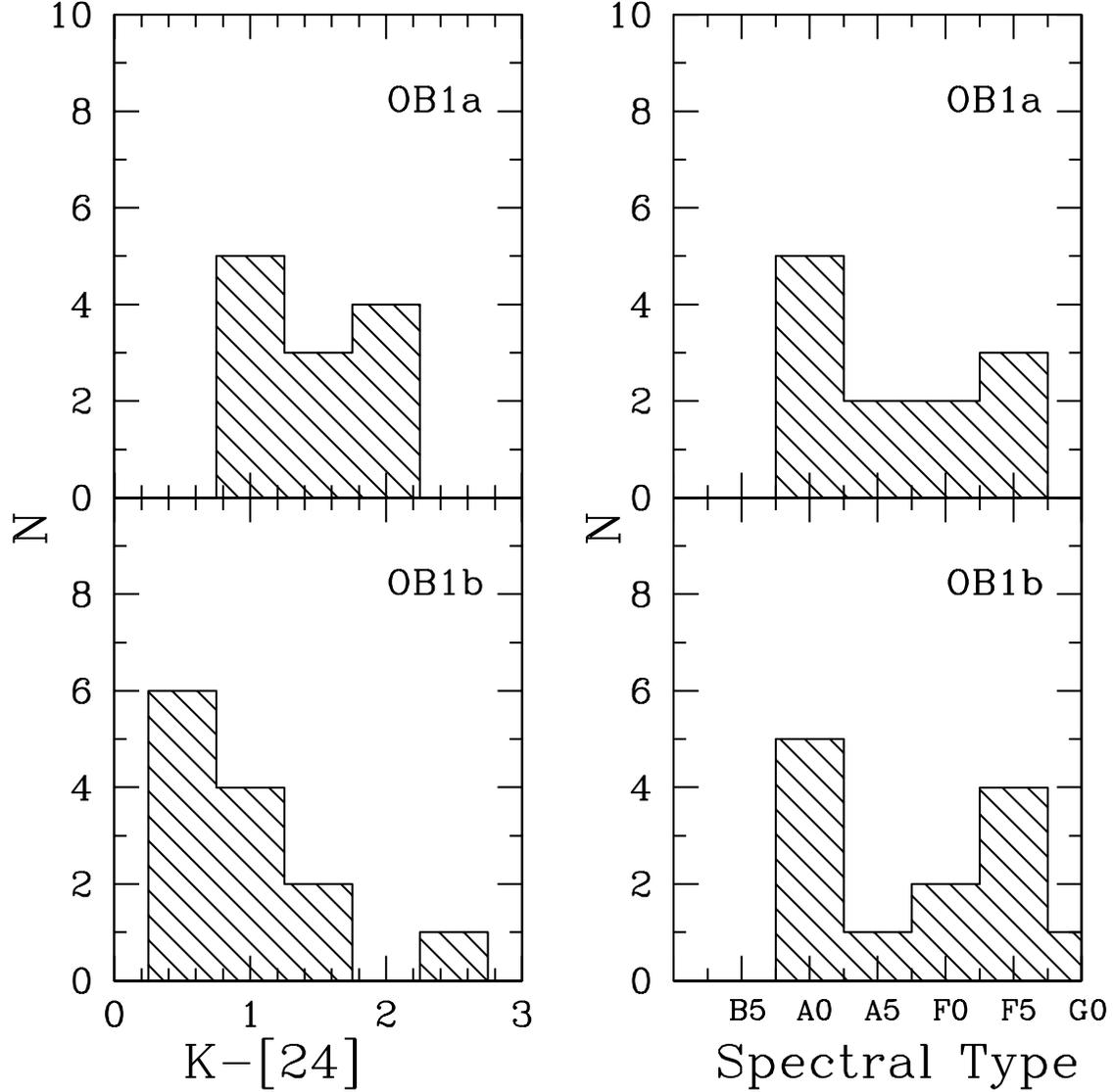}
\caption{K-[24] (left panels) and spectral type (right panels)
distributions for the 24 {\micron} excess stars in OB1a (upper panels) 
and OB1b (lower panels). 
We exclude the HAeBe stars (V346 Ori and HD290543) and the 
CBe star (25 Ori). A comparison of the distributions of spectral types
shows a Kolmogorov-Smirnov significance level of 75\%,
indicating that we are comparing similar populations. 
In contrast, the K-[24] histogram on OB1b field appears 
displaced toward lower values of K-[24] excess than the OB1a field.
A Kolmogorov-Smirnov test shows that the significance level in this case is 5\%.}
\label{fig:hist}
\end{figure}

The lower panel in Figure \ref{fig:mips} reveals that 
the stars in the debris disk region in OB1b have systematically 
less 24 {\micron} excess in comparison with OB1a; 
only two stars in OB1b, HD290543 and HD36444 (see below), have 
relative strong 24 {\micron} excess. The larger 24{\micron} excess 
in OB1a is confirmed by Figure \ref{fig:hist}, 
which shows a comparison of the distributions of the 
K-[24] colors (left panels) and the spectral types (right panels) 
for the stars in the debris disk region in 
the OB1a (upper panels) and OB1b (lower panels) fields. 
We do not include in this plot the CBe star 25 Ori.
The difference in the K-[24] histograms between OB1a and OB1b
is confirmed by a Kolmogorov-Smirnov (KS) test showing a
significance level is only 5\%. 
In contrast to the K-[24] distributions,
the spectral type distributions for both regions  
are very similar, with a significance level of 75\% in a KS test 
indicating that we are comparing similar populations.  
Also the fraction of debris disk stars candidates in 
OB1b ($\sim$38$\pm$3\%) appears to be 
slightly lower than in OB1a ($\sim$46$\pm$4\%). 
This indicates that the debris disk phenomenon appears 
to be more frequent and stronger 
in the older sub-association, OB1a.

\begin{figure}
\epsscale{0.9}
\plotone{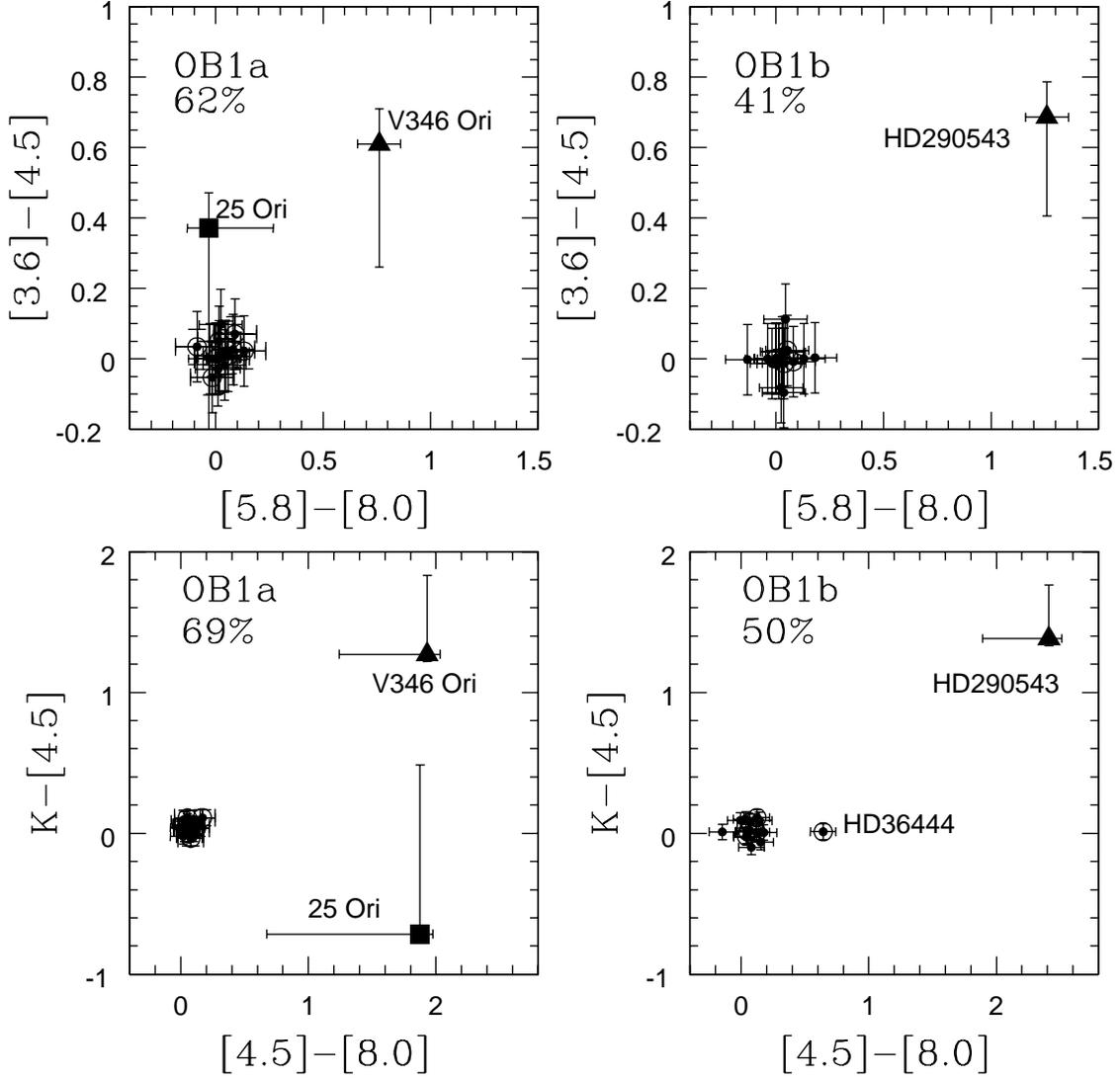}
\caption{Color-color diagrams, [5.8]-[8.0] versus [3.6]-[4.5] (upper panels) 
and [4.5]-[8.0] versus K-[4.5] (lower panels). Symbols are the same as Figures 
\ref{fig:field} and \ref{fig:mips}. The error bars for the HAeBe stars and the CBe star 
include the uncertainties from saturation effect. The IRAC excesses on the 
HAeBe stars are apparent in these diagrams.  The star HD36444 
(not included in the upper panel) has excess at 8\micron, but not at [4.5] IRAC band.
The star 25 Ori appears to have small excess in [4.5]-[8.0] color;
since the error bars reach the  photospheric region in the diagrams,
excesses in other colors are not possible to estimate. We show the percent 
of the sample plotted in each panel.}
\label{fig:irac}
\end{figure}

\begin{figure}
\epsscale{0.9}
\plotone{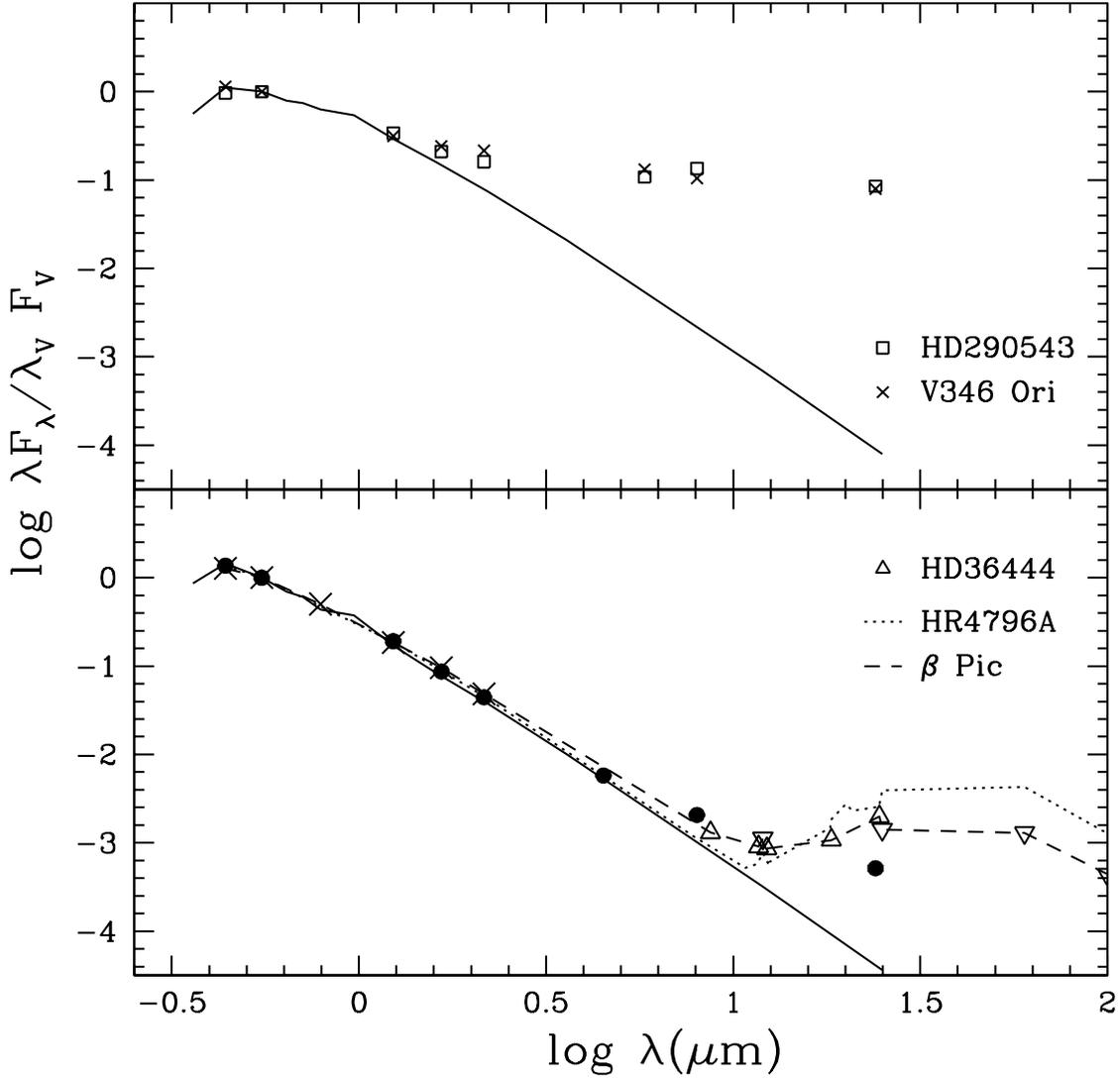}
\caption{Spectral energy distributions of HD290543 and HD36444.
The fluxes are calculated from the BV photometry \citep{hipp97}, 
2MASS photometry \citep{cutri03}, and IRAC/MIPS magnitudes.
In each case, fluxes have been corrected for reddening 
(estimated form the B-V color and the spectral type)
using the standard reddening law, and have been normalized to the V band. 
In the upper panel, SEDs of the HAeBe stars, V346 Ori (symbol X, spectral type A8) 
and HD290543 (open squares) are displayed. For comparison, 
we also show the photospheric emission \citep[solid line; ][]{kh95} of a star with a similar spectral 
type of HD294503 (A7). The lower panel shows the SEDs of the 
star HD36444 (solid circles), the star HR4796A \citep[dotted line; ][]{wahhaj05}
and the star $\beta$ Pic (dashed line). The SED of $\beta$ Pic 
was derived from magnitudes in Hipparcos and 2MASS catalogs (X's),
IRAS fluxes (inverse open triangles) and  the mid-IR data from
\citet[][open triangles]{telesco05}
The solid line represents the photospheric emission
of a star with spectral type B9.
}
\label{fig:sed}
\end{figure}

Figure \ref{fig:irac} shows the color-color diagrams, 
[5.8]-[8.0] versus [3.6]-[4.5] (upper panels) and [4.5]-[8.0] versus K-[4.5] 
(lower panels), for 
OB1a (left panels) and for OB1b (right panels).  The errors bars for
the stars 25 Ori, V346 Ori, HD290543 and HD36444 include the 
uncertainties from saturation effect, which were estimated using the
deviation from the expected photospheric colors for the brightest objects
detected in each IRAC band. Most of the stars in both sub-associations 
have photospheric IRAC colors. The HAeBe star V346 Ori has strong
excess in the IRAC bands originated in its primordial disk. 
The CBe star appears to have excess in the color [4-5]-[8.0]; 
the error bars in the others colors reach the photospheric 
regions, so it is not possible estimated excess in those colors. 
The nature of the stars HD290543 and HD36444 in OB1b 
is reported for first time in this work:

- The star HD290543 (triangle) is located in the HAeBe loci on the NIR 
color color diagram (upper-right panel in Figure \ref{fig:jhk}), 
shows H$\alpha$ emission in its spectrum (Figure \ref{fig:spec}), and 
exhibits large excess at 24{\micron} (Figure \ref{fig:mips}) and in the IRAC
bands (Figure \ref{fig:irac}), confirming it as a new HAeBe in the 
OB1b sub-association. Upper panel in Figure \ref{fig:sed} shows 
the spectral energy distribution (SED) of HD290543. For comparison, 
the SEDs of the star V346 Ori (symbol X), and the photosphere 
\citep[solid line; ][]{kh95} of a star with a similar spectral type (A7) are displayed. 

- The star HD36444 does not exhibit NIR excess in Figure \ref{fig:jhk}
and have spectral type B9 without emission lines in its spectrum
(Figure \ref{fig:spec}). This star is the only object located in the
debris disk region in Figure \ref{fig:mips} that shows excess in [8.0]
IRAC band\footnote{HD36444 is out of the IRAC fields at [3.6] and [5.8]} 
(Figure \ref{fig:irac}). The lower panel of Figure \ref{fig:sed}
shows the SED for HD36444, for comparison, the SEDs of 
the star HR4796A \citep[dotted line; ][]{wahhaj05}, 
which defines the debris disk region in Figure \ref{fig:mips},
the SED of the star $\beta$ Pic (dashed line); data from Hipparcos, 2MASS, IRAS catalogs and \citet{telesco05}, 
and  the  photosphere for A0 type star 
\citep[solid line; ][]{kh95}
are displayed. This comparison indicates that the SED of HD36444 
is similar to that of $\beta$ Pic, suggesting that these objects 
have similar evolutionary stages.

\subsection{Debris disk evolution}

Figure \ref{fig:evol} shows the excess emission at 24\micron, 
calculated as the ratio of the observed flux to that expected
from the stellar photosphere, for intermediate
mass stars in several stellar groups spanning a range of age from $\sim$2.5 Myr 
to $\sim$150 Myr. We use the color K-[24] to determine
this excess ratio (E${_{ratio}}$). At these long wavelength, the photospheric colors
are null and independent of spectral types \citep[Figure \ref{fig:mips}; see also ][]
{gorlova04, young04, rieke05}. So, for Orion OB1a and Orion OB1b sub associations (open circles),
E${_{ratio}}$=10$^{(K-[24] - 0.02)/2.5}$, where 0.02 is the mean value of K-[24] (see \S3.1).
Our definition of stars with excess (see Figure \ref{fig:mips})) agrees with that given by
\citet{rieke05}, which is represented as a solid horizontal line
(E${_{ratio}}$ =1.25). Other stellar groups included in Figure
\ref{fig:evol} are: IC 348 \citep{lada06} with age
of 2-3 Myr (star symbol),  NGC2547 \citep{young04, rieke05} with age of 25 Myr
(open triangles), M47 \citep{gorlova04, rieke05} with age of 100 Myr
(open square) and NGC2516 \citep{rieke05} with age 150
Myr (inverse open triangle).  Similarly to our objects 
in Orion, these stellar groups are within 500 pc from the sun and have been
observed using the MIPS scan mode to mapping large regions, so
together they provide an excellent data set
for statistical proposes. In addition, they have more reliable
stellar ages than individual objects in the sky. 
The dashed line represents the median value of E${_{ratio}}$ for the
stars with excess at 24 {\micron} in those stellar groups. 
By comparison, we include also stars observed by \citet{rieke05} using 
the MIPS point source mode or compiled for them from IRAS data (symbol
X). Some stars with E${_{ratio}}$$>$8 (hereafter, massive debris
disks) are represented by arrows: from left to right, HD36444 (5 Myr) 
with E${_{ratio}}$=11.8, HR4796A (8 Myr) with E${_{ratio}}$=97.2, 
$\beta$ Pic (20 Myr) with E${_{ratio}}$=21.9, and HD21362 (80 Myr) 
with E${_{ratio}}$=8.4. These observations have been compared with
the results of two collisional cascade models for A type stars 
\citep[][ hereafter KB models]{kb04b,kb05}:  an ``outer model'' for the
region from 30 AU to 150 AU (dot-dashed line), where the icy planets
are formed, and an ``inner model'' for the region from 3 AU to 20 AU
(dotted line), where rocky planets are expected to grow.

The outer KB model predicts that icy objects ($\sim$1000-3000 Km) begin to form
at the inner edge of the model at $\sim$10-20 Myr. These objects 
stir up leftover planetesimals along their orbits and 
initiate a collisional cascade, in which smaller planetesimals 
($\sim$1-10 Km) are slowly ground into the fine dust observed in debris disks 
around young stars. 
The luminosity of the 
debris disk declines exponentially with stellar age, since the reservoir of smaller 
objects diminishes and the products from the collisions are removed mainly by 
radiation pressure and Poynting-Robertson drag \citep[e.g., ][]{dominik03, chen06}. 
Since the KB models do not include the sublimation of ices, the excesses
observed in the inner model could be overestimated. 
In particular, grains of water ice (${\rm H_2O}$)  begin to sublimate at $\sim$120 K \citep{fraser01}; 
assuming that the dust particles act like a black body in radiative equilibrium \citep{chen06}, for an A0 star
(T$\rm{_{eff}}$=9760 K, R=2.4 R$\sun$)  this temperature is
reached at $\sim$37 AU from the star; the sublimation is completed at
18 AU where the grain temperature is $\sim$170 K. 
The central clearing  observed in some debris disks can be 
explained if the ice sublimation
is an important mechanism for dust removal \citep{chen06, decin03}. 
The KB models have been  scaled (scale factor$\sim$0.5) and 
shifted ($\sim$0.3 in log time units) in Figure \ref{fig:evol}, so the peak in the outer model 
agrees with the peak observed in debris disk at 10 Myr (Orion OB1a). 
The vertical scale factor can be explained if: a) the dust production per
collision is less than the model's prediction (e.g., interactions between solid bodies
and the gas damp particle velocities and remove particles from the model grid); or/and
b) icy sublimation diminishes the dust produced at 30-37 AU from
the star.  Since the initial conditions in the KB models include objects 
with radius $\sim$1-1000 m \citep{kb04b}, the zero point for the 
age in the KB models could be different to that from the stellar evolution studies \citep[e.g., ][]{palla93}. 
However, since the growth time to reach objects with diameters of a few meters 
in the disk is relatively short \citep[$\lesssim$10$^6$ yr; ][]{weidenschilling97, dullemond05},  
this zero point difference is not significant for stars
with few Myr or older ( Kenyon 2006, private communication). 
Besides the zero point differences in age, the shift in the age of the KB
models could be explained if:
a) the timescales from KB models are $\sim$30\% to $\sim$40\% longer than
the actual evolution times, due the radial resolution in the grids
\citep{kb04b}; and/or b) cascade collisions occur in the region between the inner and outer
model (20-30AU), where the 1000 Km objects could be formed faster than in the outer model 
\citep[see ][ for more details about uncertainties and limits of the models]{kb02, kb04b,kb05}.

In general terms, observations of debris disks follow the trend predicted by the KB
models. The upper envelope in excess ratio in debris disks with age $\gtrsim$10 Myr can be
fitted by the KB outer model.  In Orion OB1b, at 5 Myr old, the
outer KB model does not predict enough excess emission to explain the observed
excesses. A small dust contribution from collisional cascade 
in the inner disk ($\lesssim$20-30AU) or remaining primordial dust   
in the disk could explain the differences. In IC 348 ($\sim$2.5 Myr) 
statistical fluctuations due to the small size of the sample 
could be important; on the other hand,
the relative large excess could be explained if there was a significant amount of
remaining primordial dust in the disk, and these objects were in an intermediate 
phase between HAeBe and debris disks.  

The stars with massive debris disks  could be explained if the 
observed dust is  produced by eventual collisions between
2 large objects \citep[$\gtrsim$1000 Km; ][]{kb04b,kb05}. 
However, an alternative explanation  for younger stars
($\lesssim$20Myr) is that those objects are in a phase
between HAeBe star and true debris systems \citep{rieke05} retaining 
some of the primordial dust.
In any case, our observations confirm that this intermediate phase 
has to be very brief; if the star HD36444 is in this phase, 
the fraction at 5 Myr (OB1b) is $\sim$3\%, and null if 
the 24{\micron} excess in HD36444 is from a massive debris disk. 
In contrast to the small fraction of primordial disk 
reported in \citet{hernandez05} for the association OB1 (3-5\%),
the total disk fractions (primordial disks + debris disks) are  
50\% and 40\% for OB1a and OB1b, respectively. Since some objects 
could have a cooler disks, which are not detected at 24\micron, 
these fractions are inferior limits.  So, most of the stars 
in the mass range of HAeBe are born surrounded by disks, 
which evolve to debris disks in few Myr.

\section{Conclusions}
\label{sec:con}
We have combined Spitzer IRAC and MIPS observations, 2MASS data,
and optical spectra to study the 
properties of the disks around early type stars (B, A, F)
in fields of two sub-associations of Orion OB1: 
OB1a with an age of 10 Myr \citep{briceno06} and OB1b with 
an age of 5 Myr \citep{briceno05}.
In each field we find only one star with an optically
thick disk which identifies them as a HAeBe star;
of these, V346 Ori in Ori OB1a was already known \citep[e.g., ][]{hernandez05, vieira03, the94},
but HD290543 in Ori OB1b is reported here for the first time.
The star HD36444 in OB1b is the only non-emission line star in our
sample with excess at the IRAC [8.0] and the MIPS [24] 
bands. The SED of this object is similar to that of $\beta$ Pic; 
both objects have a relatively large 24 {\micron} excess 
that could be explained if they have massive debris disks 
produced by the eventual collisions between two large ($>$1000 Km) objects
\citep{kb05, kb04b}. 
An alternative explanation could be that these objects are in a
intermediate phase between HAeBe and true debris systems \citep{rieke05}. 
The small or null fraction of objects in this phase found in our samples 
confirms that the lifetime of this intermediate phase is very brief. 

We find a population of early type stars without excess in the IRAC
bands and without emission lines in their optical spectra, but with
varying degrees of K-[24] excess. We identify these stars as debris
disks. We find that in the older  OB1a sub-association, the debris 
disk phenomenon, diagnosed by the mid-IR excess, 
is stronger and more frequent than in the younger 
sub-association OB1b.

We have put together our observations with those
of other stellar groups of different ages, and compared
them with the theoretical models of \citet{kb05}.
We find a peak in the debris disk phenomenon at 10 Myr
indicated by our observations of Ori OB1a in agreement
with theoretical predictions, in which the peak is 
associated to the formation of large icy objects (1000 Km)
in 10-20 Myr, which stir up the smaller objects in the
disk and produce a collisional cascade, in which $\sim$1-10 Km
planetesimals are converter in fine dust grains \citep{kb04b, kb05}.
Stellar groups older that 10 Myr follow the  predictions of 
collisional cascades in the outer model (30-150 AU,  
where the icy planets are formed) proposed by \citet{kb05} 
for A type star.
The relative small excess and small fraction of debris disks
observed at 5 Myr (Ori OB1b) could be associated 
to a phase between the clearing of the primordial disk, 
$<$3Myr \citep{hernandez05}, and the formation of large icy 
objects, 10-20Myr \citep{kb04b}. In this phase, small 
amount of remaining primordial dust and small production 
of second generation dust are expected. On the other hand,
these amounts of dust (remaining primordial dust 
and/or second generation dust)  could explain the 
differences between our observations at 5 Myr (OB1b) and the 
outer model of \citet{kb05}. 
The large excesses in the younger stellar group IC 348 ($\sim$2.5 Myr)
could indicate that a significant amount of primordial dust
remains in the disk. However,  small statistical number fluctuation
can affect the excesses observed and a more detailed study is necessary to 
establish the presence of primordial dust in these systems.

\section{Acknowledgments}
\label{sec:graci}

\begin{figure}
\epsscale{0.7}
\plotone{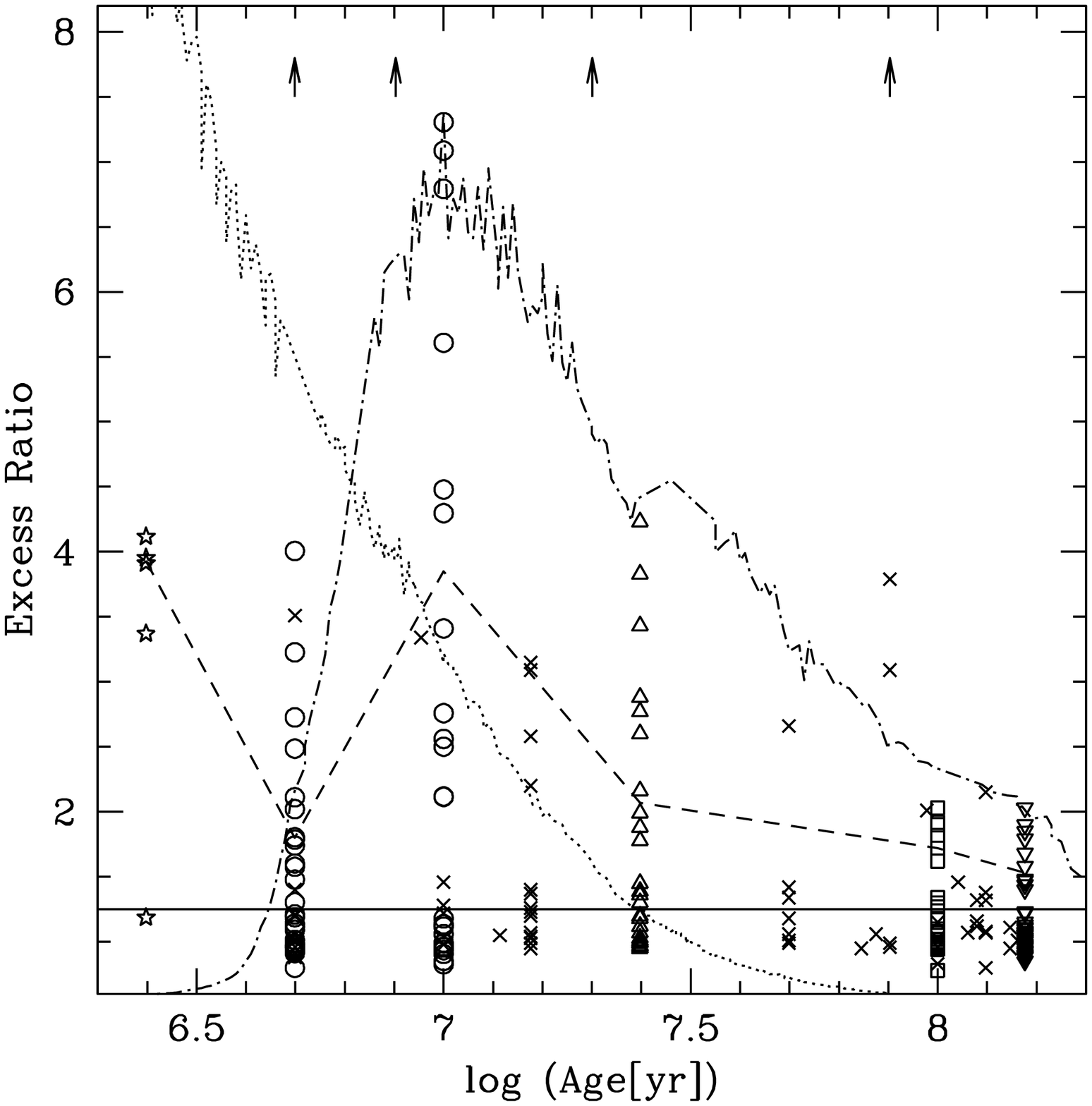}
\caption{ 24 {\micron} excess versus logarithm of the age. The excess
ratio is calculated as the ratio of the observed flux at 24 {\micron} to that
expected  from the stellar photosphere. Our observations in OB1
association (open circles) are plotted with other stellar groups
mapped with MIPS: IC 348 \citep[star symbols, $\sim$2.5 Myr; ][]{lada06},  
NGC2547 \citep[open triangles, 25 Myr; ]{young04, rieke05}, 
M47 \citep[open square, 100 Myr; ][]{gorlova04, rieke05} and
NGC2516 \citep[inverse open triangles, 150Myr; ][]{rieke05}. 
The solid line represents the definition of excess stars from \citet{rieke05},
which agrees with our limit for debris disks (see Figure \ref{fig:mips}).
The dashed line represents the median values for debris disk objects in each
stellar group. The dotted line and the dot-dashed line represent 
the inner (3-20 AU) and the outer (30-150 AU)  collisional cascade 
model from \citet{kb05}, respectively. 
The stellar groups with age $\gtrsim$10 follow the trend
expected  by the outer collisional cascade model. 
At 5 Myr collisional cascade in inner region (20-30 AU) or remain
primordial dusts could explain the differences between the observation
and the collisional cascade model at 30-150 AU. Since the models do 
not include ice sublimation process, the excess predicted by the 
inner model could be overestimated if the grains are icy in the 
inner region of the disk. In this case, a significant amount of 
primordial dust can be expected at IC 348. By comparison, we include 
stars observed by \citet{rieke05} using the MIPS point source mode 
or compiled for them from IRAS data (symbol X). The upward-pointing 
arrows are objects with a possible massive debris disks; from left to right, 
HD36444 (in Ori OB1b), HR 4796A, $\beta$ Pic, and HD21362.
}
\label{fig:evol}
\end{figure}

We thank Scott Kenyon for useful comments and for providing 
us the models plotted in Figure \ref{fig:evol},
Francesco Palla for providing the isochrones of PMS stars, 
and Massimo Marengo for insightful communications. We also thank 
Perry Berlind and Susan Torkarz of the SAO Telescope Data Center for carrying out the
observation and data reduction of the spectra. This 
publication make use of data products from the Two Micron All Sky 
Survey, which is a joint project of the University of Massachusetts
and the Infrared Processing and Analysis Center/California Institute 
of Technology. This work is based on observations
made with the Spitzer Space Telescope (GO-1 3437), which is operated by the
Jet Propulsion Laboratory, California Institute of Technology under
a contract with NASA. Support for this work was
provided by NASA through an award issued by JPL/Caltech.
This publication was supported in part by the NASA grants 
NAG5-9670 and NAG10545, NSF grant AST-9987367 and grant No. 
S1-2001001144 of FONACIT, Venezuela.

\begin{deluxetable}{ccccccc}
\tabletypesize{\scriptsize}
\tablewidth{0pt}
\tablecaption{MIPS fields\label{tab:mips}}
\tablehead{
\colhead{Name} & \colhead{D$_{Ref}$} & \colhead{Age} & \colhead{$\alpha_C$} & \colhead{$\delta_C$} & \colhead{Area} & \colhead{\av(ref)}\\
\colhead{ } & \colhead{ pc } & \colhead{Myr} & \colhead{Deg} & \colhead{Deg} & \colhead{Deg$^2$} & \colhead{mag $^1$}
}
\startdata
Ori OB1a      & 335$\pm$13 & 10  & 81.2986 & 1.6438 &  1.10 & 0.12 \\
Ori OB1b      & 443$\pm$18 & 5   & 82.7585 & -1.7050 & 1.42 & 0.60 \\
\enddata
\tablenotetext{1}{ 90\% of the field have \av$<$\av(ref)}
\tablenotetext{-}{ Distances from \citet{hernandez05}}
\tablenotetext{-}{ Ages from \citet{briceno05,briceno06}}
\end{deluxetable}

\begin{deluxetable}{cccccccccccc}
\tabletypesize{\scriptsize}
\tablecolumns{12}
\tablewidth{0pt}
\tablecaption{Intermediate mass stars in OB1a and OB1b\label{tab:obs} }
\tablehead{
\colhead{2MASS} & \colhead{Name} & \colhead{[3.6]} & \colhead{[4.5]} & \colhead{[5.8]} & \colhead{[8.0]} & \colhead{[24]} & \colhead{F$_{24\micron}$} & \colhead{Excess} & \colhead{Spectral} & \colhead{ref} & \colhead{Disk} \\
\colhead{     } & \colhead{    } & \colhead{mag}   & \colhead{mag}   & \colhead{mag}   & \colhead{mag}   & \colhead{mag}  & \colhead{mJy}        &\colhead{ratio} & \colhead{type}     &  \colhead{  } & \colhead{ }
}
\startdata
\cutinhead{Orion OB1a}
 05224795+0143002 &   HD35150 &   9.43 &   9.41 &   9.37 &   9.24 &   7.64 &    6.39 &    5.61 &    A0 &  1 &    debris \\ 
 05230192+0141489 &   HD35177 &   s  &   s  &   8.37 &   8.36 &   8.40 &    3.18 &    0.99 &    B8 &  2 &    no-disk\\ 
 05230686+0118237 &  HD287787 &   9.96 & \nodata &   9.59 & \nodata &   8.66 &    2.50 &    3.41 &    A7 &  1 &    debris \\ 
 05231014+0108225 &   HD35203 & \nodata & \nodata & \nodata & \nodata &   8.25 &    3.67 &    0.91 &    B7 &  2 &    no-disk\\ 
 05234591+0150334 &  HD287773 &   9.48 &   9.50 &   9.51 &   9.47 &   9.63 &    1.02 &    0.94 &    A2 &  4 &    no-disk\\ 
 05235020+0132529 &  HD287845 &   9.46 &   9.46 &   9.40 &   9.38 &   9.39 &    1.28 &    1.14 &    F5 &  4 &    no-disk\\ 
 05240786+0138000 &   HD35332 &   9.38 &   9.36 &   9.32 &   9.24 &   7.24 &    9.32 &    7.30 &    B9 &  1 &    debris \\ 
 05241272+0134121 &  HD287846 &   9.82 &   9.88 &   9.79 &   9.81 &   7.79 &    5.58 &    6.79 &    F4 &  1 &    debris \\ 
 05241300+0146421 &   HD35351 &   9.16 &   9.09 &   9.10 &   9.01 &   8.24 &    3.68 &    2.11 &    A0 &  1 &    debris \\ 
 05241392+0115432 &  HD287860 & \nodata & \nodata & \nodata & \nodata &   8.24 &    3.68 &    2.76 &    F6 &  1 &    debris \\ 
 05241447+0151479 &   \nodata &   9.86 &   9.89 &   9.84 &   9.83 &   9.99 &    0.74 &    0.96 & \nodata &  \nodata &    no-disk\\ 
 05242074+0135266 &   HD35367 &   9.58 &   9.57 &   9.58 &   9.52 &   8.03 &    4.47 &    4.30 &    A1 &  1 &    debris \\ 
 05242421+0141333 &  HD287842 &   9.35 &   9.34 &   9.36 &   9.31 &   9.37 &    1.30 &    1.06 &    A3 &  4 &    no-disk\\ 
 05243861+0148388 &   HD35409 &   s  &   s  &   8.44 &   8.43 &   8.29 &    3.53 &    1.18 &     A &  3 &    no-disk\\ 
 05244482+0150472 &   25 Ori &    s  &   s  &   s &    s  &   3.52 &  284.12 &    5.40 &    B1 &  1 &   CBe \\ 
 05245456+0158083 &  HD287838 & \nodata & 9.35 & \nodata &   9.25 &   9.44 &    1.22 &    0.96 &    A2 &  4 &    no-disk\\ 
 05251029+0115314 &  HD287861 & \nodata & \nodata & \nodata & \nodata &   7.47 &    7.48 &    7.09 &    A3 &  1 &    debris \\ 
 05251139+0155242 &   HD35501 &   s    &   s  &   7.62 &   7.61 &   7.78 &    5.64 &    0.83 &    B8 &  2 &    no-disk\\ 
 05251164+0133296 &  HD287847 &   9.52 &   9.51 &   9.47 &   9.46 &   9.47 &    1.19 &    1.12 &    A3 &  4 &    no-disk\\ 
 05253978+0138183 &  HD287850 &   9.67 &   9.64 &   9.50 &   9.59 &   8.93 &    1.95 &    2.12 &    A9 &  1 &    debris \\ 
 05254865+0123220 &  HD287854 &   9.73 & \nodata &   9.68 & \nodata &   8.16 &    3.99 &    4.48 &    F0 &  1 &    debris \\ 
 05260368+0148293 &  HD287851 &   9.83 &   9.83 &   9.79 &   9.75 &   8.87 &    2.07 &    2.56 &    F3 &  1 &    debris \\ 
 05261198+0153357 &   HD35625 &   9.33 &   9.28 &   9.25 &   9.24 &   8.32 &    3.42 &    2.50 &    A0 &  1 &    debris \\ 
 05264810+0204058 &   HD35716 &   s  &  8.68 & \nodata &   8.66 &   8.83 &    2.14 &    0.85 &    B9 &  2 &    no-disk\\ 
 05271919+0136224 &   HD35791 &   9.00 &   8.90 &   8.91 &   8.89 &   8.95 &    1.92 &    0.99 &    B9 &  5 &    no-disk\\ 
 05244279+0143482 &  V346 Ori &   s  &   s  &   6.12 &   5.36 &   2.12 & 1034.41 &  376.36 &    A9 &  1 &  primordial \\ 
\cutinhead{Orion OB1b}
 05285515-0145387 &   HD36057 & \nodata& \nodata& \nodata& \nodata&   8.28 &    3.57 &    1.09 &    A0 &  3 &    no-disk\\ 
 05291353-0147463 &  HD290534 &   9.48 &   9.46 &   9.46 &   9.40 &   8.38 &    3.25 &    2.73 &    F2 &  1 &    debris \\ 
 05291785-0143391 &  HD290533 &   9.53 &   9.54 &   9.49 &   9.48 &   9.64 &    1.01 &    0.92 &    F8 &  4 &    no-disk\\ 
 05292120-0200309 &   HD36118 &  s &  s &   8.54 &   8.58 &   8.73 &    2.34 &    0.94 &    A0 &  3 &    no-disk\\ 
 05292269-0131537 &  HD290530 &   9.37 &   9.28 &  \nodata &   9.16 &   8.97 &    1.89 &    1.48 &    F8 &  1 &    debris \\ 
 05292958-0215399 &  HD294152 & \nodata& \nodata& \nodata& \nodata&   8.70 &    2.43 &    2.49 &    F4 &  1 &    debris \\ 
 05294888-0139281 &   HD36176 &  s &  s &   8.78 &   8.63 &   8.86 &    2.09 &    0.93 &    A0 &  3 &    no-disk\\ 
 05295026-0102088 &  HD290521 & \nodata& \nodata& \nodata& \nodata&   8.78 &    2.24 &    1.80 &    A6 &  1 &    debris \\ 
 05295547-0215163 &   HD36185 & s & \nodata&   8.70 & \nodata&   8.67 &    2.48 &    1.01 &    A0 &  3 &    no-disk\\ 
 05300287-0208595 &  HD290548 &  10.02 & \nodata&   9.94 & \nodata&   9.48 &    1.18 &    1.74 &    F6 &  1 &    debris \\ 
 05300439-0144584 &   HD36219 & s & s &   7.80 &   7.78 &   7.91 &    4.99 &    0.91 &    B8 &  2 &    no-disk\\ 
 05301868-0201575 &  HD290543 & s & s &   6.42 &   5.16 &   2.13 & 1027.77 &  537.53 &    A7 &  1 &    primordial \\ 
 05303822-0139029 &  HD290536 &   9.54 &   9.53 &   9.51 &   9.49 &   9.64 &    1.02 &    0.99 &    F2 &  4 &    no-disk\\ 
 05305984-0202024 &  HD290541 &   9.53 &   9.53 &   9.54 &   9.68 &   9.78 &    0.89 &    0.80 &    F0 &  4 &    no-disk\\ 
 05312070-0158522 &   HD36393 &  s & s &   8.68 &   8.61 &   8.48 &    2.96 &    1.19 &    B8 &  3 &    no-disk\\ 
 05312120-0205569 &   HD36394 & s &  s &   7.96 &   7.96 &   7.93 &    4.90 &    1.60 &    B9 &  1 &    debris \\ 
 05313136-0149332 &  HD290540 &   9.50 &   9.51 &   9.47 &   9.39 &   8.75 &    2.31 &    2.02 &    A0 &  1 &    debris \\ 
 05314048-0107332 &   HD36444 & \nodata&   8.98 & \nodata&   8.33 &   6.31 &   21.93 &   11.80 &    B9 &  1 &    debris $^{a}$ \\ 
 05314469-0130582 &   \nodata &   9.92 &   9.87 & \nodata&   9.80 &   9.14 &    1.61 &    2.11 &    F7 &  1 &    debris \\ 
 05314814-0124566 &  HD290594 &   9.69 &   9.67 &   9.64 &   9.61 &   9.75 &    0.92 &    0.95 &    F5 &  4 &    no-disk\\ 
 05315101-0159448 &   \nodata &   9.80 &   9.80 &   9.77 &   9.81 &   9.78 &    0.90 &    1.11 & \nodata & \nodata &    no-disk\\ 
 05320405-0128445 &   HD36502 &   9.22 &   9.23 &   9.13 &   9.15 &   9.18 &    1.55 &    0.96 &    B9 &  3 &    no-disk\\ 
 05320525-0137136 &  HD290598 &   9.69 &   9.70 &   9.70 &   9.66 &   9.05 &    1.75 &    1.79 &    A0 &  1 &    debris \\ 
 05321963-0124133 &  HD290592 &   9.33 &   9.41 &   9.28 &   9.26 &   9.44 &    1.23 &    0.92 &    G0 &  4 &    no-disk\\ 
 05322179-0101191 &  HD290585 & \nodata& \nodata& \nodata& \nodata&   8.92 &    1.97 &    3.23 &    F1 &  1 &    debris \\ 
 05324082-0148341 &  HD290610 &   9.58 &   9.67 &   9.57 &   9.53 &   9.55 &    1.10 &    1.21 &    A0 &  4 &    no-disk\\ 
 05324134-0135305 &   HD36591 &  s &  s &   6.00 &   5.87 &   6.31 &   21.91 &    0.68 &    B1 &  2 &    no-disk\\ 
 05324164-0202131 &   HD36592 &   9.02 &   8.91 &   8.91 &   8.87 &   8.60 &    2.66 &    1.30 &    B9 &  3 &    no-disk\\ 
 05324975-0211494 &   HD36617 & s & \nodata&   8.39 & \nodata&   8.18 &    3.92 &    1.14 &    A0 &  2 &    no-disk\\ 
 05325042-0136026 &  HD290599 &   9.60 &   9.59 &   9.60 &   9.41 &   9.93 &    0.78 &    0.74 &    A2 &  4 &    no-disk\\ 
 05330558-0143155 &  HD290609 & \nodata& \nodata& \nodata& \nodata&   6.96 &   11.96 &    4.01 &    B9 &  1 &    debris \\ 
 05330734-0143021 &   HD36646 & \nodata& \nodata& \nodata& \nodata&   6.83 &   13.49 &    0.97 &    B3 &  2 &    no-disk\\ 
 05331305-0203129 &   \nodata & \nodata& \nodata& \nodata& \nodata&   9.97 &    0.75 &    1.58 &    F5 &  1 &    debris \\ 
 05331540-0143124 &  HD290608 & \nodata& \nodata& \nodata& \nodata&   9.05 &    1.76 &    1.01 &    B8 &  3 &    no-disk\\ 
\enddata
\tablenotetext{a}{ Massive debris disk}
\tablenotetext{s}{ Saturated}
\tablecomments{Spectral type reference: 1 This work; 2 \citet{hernandez05}; 3 \citet{kharchenko01}; 4 \citet{nesterov95}; 5 \citet{schmidt93}}
\end{deluxetable}

\end{document}